\documentstyle[preprint,prb,aps]{revtex}
\preprint{HKBU-CNS-9820}
\begin{document}
\draft
\title{Wave transmission, phonon localization and heat conduction of 1D 
Frenkel-Kontorova  chain\footnote{to appear in Phys. Rev. B {\bf 
59}, B1 No. 13 (1999).}}

\author{Peiqing Tong$^{1,2,3}$, Baowen Li$^1$, and Bambi 
Hu$^{1,4}$} 
\address{
$^{1}$Department of Physics and Centre for Nonlinear Studies, Hong Kong 
Baptist University, Hong Kong, China,\\ 
$^{2}$CCAST(WORLD LABORATORY) P.O. Box 8730, Beijing, 100080, China\\
$^{3}$Department of Physics, Nanjing Normal University, Nanjing, Jiangsu, 
210097, China\\
$^{4}$Department of Physics, University of Houston, Houston, Texas 77204, USA}

\date{\today}
\maketitle
\begin{abstract}
We study the transmission coefficient of a plane wave through a 1D finite 
quasi-periodic system -- the Frenkel-Kontorova (FK) model --  embedding 
in an infinite uniform harmonic chain.
By varying the mass of atoms in the infinite uniform chain, we obtain 
the transmission coefficients for {\it all} eigenfrequencies. The phonon 
localization of the incommensurated FK chain is
also studied in terms of the transmission 
coefficients and the Thouless exponents. Moreover, the heat conduction of 
Rubin-Greer-like model for FK chain at low temperature is calculated.
It is found that the stationary heat flux $J(N)\sim N^{\alpha}$, and
$\alpha$ depends  on the strength of the external potential. 
\end{abstract}


\section{introduction}
\label{sec:one}

In recent years, there has been growing interest in studying
incommensurate structures and commensurate-incommensurate phase
transitions in condensed matter physics and dynamical systems
\cite{fr38,ba82,au831,pe83,au832,co83,ch86,au91,ma91,sh92,li92,ch79,gr79,sh82,bu96}.
On the one hand, incommensurate structures appear in many physical systems 
such as
quasi-crystals, two-dimensional electron systems, magnetic superlattices,
charge-density waves, organic conductors, and various atomic monolayers
adsorbed on crystalline substrates.  One of the simplest prototypes 
for incommensurate structures is the Frenkel-Kontorova (FK)
model\cite{fr38} which describes a one-dimensional (1D)
chain of atoms with an elastic nearest neighbor interaction, and subjected
to an external periodic potential.  If the mean distance between
consecutive atoms is not in rational ratio to the period
of the
external potential, the corresponding state is incommensurate. 
It has been shown by Aubry\cite{au831,pe83,au832} in his pioneering work that
there exists two different configurations 
for an incommensurate state by changing the strength of the external 
potential. These two configurations are called the sliding phase and 
pinned phase, and the 
 transformation between these two phases has been called
the {\it transition by breaking of analyticity}.
On the other hand, the ground state equation of the 1D FK model 
is nothing but the standard map which is widely studied in nonlinear 
dynamical theory\cite{ch79,gr79,sh82}.
The two incommensurate configurations 
correspond to the invariance circle and the Cantorus, respectively.

In the past two decades, most of the works about 1D FK model have been 
focused on the
study of the ground state. Little attention has been
paid to the linear excitation or phonon modes. 
As is well known that the excitation of phonons is very important, it 
determines the fundamental properties, such as the wave transmission, 
heat 
conduction and other low temperature thermodynamics properties, of the 
underlying material. In turn, the study of 
the transmission coefficients and energy transport gives us insights of 
the excitation. In this paper we shall study systematically these related 
properties, namely, the wave transmission, the heat conduction and the 
phonon 
localization in 1D FK model. Through this study we shall have a clear 
picture about how the macroscopic phenomena such as the transmission 
coefficients and heat conduction are related to the localization 
properties of the phonon excitation. Moreover, our study suggests a possible 
way 
for experimental observation of the phonon gap which characterize the 
{\it phase transition by breaking analyticity}.

Recently, Burkov {\it et
al.}\cite{bu96} have studied the localization properties of the phonon
eigenstates. They solved the equation of phonon numerically for a
FK chain  of finite length. It is  found that the  
phonon eigenstates are
extended and quasiperiodic functions for $V<V_c$, whereas for $V>V_c$ 
the eigenstates at band edges of phonon spectrum are more 
localized than that one in the middle of the spectrum, but no exponential 
localization
states have been found\cite{bu96}.  In their numerical study, the
localization property is inferred by computing the participation ratio.
Ketoja and
Satija\cite{ke97} have studied the eigenfunctions corresponding to the 
minimum frequency $\omega_{min}$ of the phonon spectra in Cantorus regime by 
using an
exact decimation scheme. The phonon eigenstates of the 
minimum frequency are found to be critical for $V>V_c$. 
 
The method used by Burkov {\it et al} is limited to the small 
system size. In our calculation we shall study the wave transmission by 
making use of the transfer matrix method due to the following reason.
As is known that there is a nice correspondence between the phonon and
electron properties. The transmission coefficient has been found to be 
a very useful quantity for the study of localization of electron
eigenstates \cite{li86}. The transfer matrix method 
is a powerful method to study the systems with much larger size. In this 
paper, we shall study systematically the
properties of the phonon eigenstates by computing the transmission 
coefficients of
a plane wave through a 1D FK chain. By using the transmission coefficient, we
are not restricted to only one frequency, instead we can study the
localization properties for {\it all} frequencies. To quantify the 
localization, we shall also calculate the Thouless exponents in addition 
to the conventional participation ratio and transmission coefficients. 

Another interesting and fundamental property that is related to the phonon
excitation is the thermal conductivity. Recently, there has been a renascence
interest in heat conduction in a variety of 1D
systems\cite{ca84,pr92,le97,hu98,fi98,ph98,al98}, since this problem is
essential for our understanding of the microscopic origin of the
macroscopic irreversibility\cite{Lebowitz98}.  At low temperature, the
linear excitations of the underlying system are most important for heat 
conduction. 
Rubin and Greer\cite{ru71,oc74} have established the relation between the
stationary energy flux and the transmission coefficients of the phonons. They
found that the stationary energy flux approach to a finite positive value
as the number of atoms goes to infinity for uniform periodic harmonic
chains, and the thermal conductivity is proportional to $N$. 
However, 
for random mass binary harmonic chains,
the stationary energy flux is
found to be proportional to $N^{-\frac{1}{2}}$ as the number of atoms $N$ 
tends to infinity, and the thermal conductivity is thus proportional to
$N^{\frac{1}{2}}$. Therefore, how the stationary energy flux and the
thermal conductivity depends on the particle number $N$ for an
incommensurate system is of great interest. This will be also 
investigated in the present paper.

The paper is organized as the follows. In Sec. \ref{sec:two}, we shall 
describe the model and numerical method for 
calculating the wave transmission coefficient. The phonon localization and 
heat conduction problems shall be discussed in Sec \ref{sec:three} and Sec. 
\ref{sec:four}, respectively. A brief discussion and conclusion is 
given in Sec. \ref{sec:five}.

\section{The model and numerical method}
\label{sec:two}

The 1D FK model can be described by
\begin{equation}
H=\sum_{n} \left[ \frac{p_n^2}{2m}+{\frac{1}{2}(x_{n+1}-x_{n}-a)^2-V\cos 
(x_n)} \right].
\label{eq:ham}
\end{equation}
where $p_n$ and $x_n$ are the momentum and position of the $n$th atom, 
respectively. $V$ is the strength of the external potential.  $a$ is the 
distance between consecutive atoms without external potential. 
Aubry and LeDaeron\cite{au832} showed that the minimum energy 
configurations are periodic when $a/2\pi$ is rational (commensurate model)
and quasiperiodic when $a/2\pi$ is irrational (incommensurate model).
For an incommensurate FK model, there are qualitatively different ground 
state configurations separated by {\it the transition by breaking 
analyticity} predicted by Aubry. For each irrational $a$ there exists a 
critical value $V_c$ of the external potential. The  
$V_c=0.9716354\cdots$\cite{gr79,sh82} corresponds to the most irrational 
number, golden mean value $a/2\pi=\frac{\sqrt{5}-1}{2}$.
Without loss of generality, we restrict ourselves to this particular 
value of $a$ in the numerical calculations throughout the paper.

The phonon equation describing the physical stability of the atoms in the 
FK model at ground state is 
\begin{equation}
\psi_{n+1}+\psi_{n-1}-[2+V\cos(x_n^0)]\psi_n=-\omega^2\psi_n,
\label{eq:phon}
\end{equation} 
where$x_n^0$ is the equilibrium position of the $n$th atom in ground state, 
$\psi_n$ is displacement of the $n$th atom from its equilibrium position and
$\omega$ is the  eigenfrequency.

This equation can be written  in the form of transfer matrix
\begin{equation}
\left( \begin{array}{c}
\psi_{n+1}\\
\psi_{n}\\
\end{array}\right)=
T_{n}\left(\begin{array}{c}
\psi_n\\
\psi_{n-1}\\
\end{array}\right)
\end{equation}
with
\begin{equation}
T_n=
\left( \begin{array}{cc}
-\omega^2+2+V\cos(x_n^0) & -1\\
1 & 0\\
\end{array}\right).
\end{equation}
Thus
\begin{equation}
\begin{array}{lllll}
\left(\begin{array}{c}
\psi_{N+1}\\
\psi_{N}\\
\end{array}\right)=
P_N\left(\begin{array}{c}
\psi_1\\
\psi_0\\
\end{array}\right)
& & \mbox{with} & &
P_N=T_NT_{N-1}\cdots T_1.\\
\end{array}
\label{eq:matrix}
\end{equation}

In order to study the transmission of a plane wave through 
the 1D FK system, we first consider a uniform harmonic atom chain with 
atom mass $m_0$, and the external potential $V$ is equal to zero. 
The eigenstates and eigenfrequencies of this chain are simply
\begin{equation}
\begin{array}{lllll}
\psi_n=A_+e^{iqna}+A_-e^{-iqna} & & \mbox{and} & &
\omega^2=\frac{4}{m_0}\sin^2(\frac{1}{2}qa),\\
\end{array}
\end{equation}
respectively. 

Then we replace the segment between $n=1$ and $n=N$ by a finite 
incommensurate FK chain. The atom in the FK chain has mass $m$.
Now we consider an incoming plane wave from $n=-\infty$ with frequency 
$\omega=\sqrt{\frac{4}{m_0}}\sin(\frac{1}{2}qa)$. Thus
in the range of $n\geq N+1$ there is 
only outgoing wave. That is
\begin{equation}
\begin{array}{lll}
\psi_n=Ae^{iqna}+A_{R}e^{-iqna} & \mbox{for} & n\leq 0 \\
\psi_n=Be^{iqna} & \mbox{for} & n\geq N+1\\
\end{array}
\label{eq:wave}
\end{equation}
From Eqs. (\ref{eq:matrix}) and (\ref{eq:wave}), after long calculation, we 
obtain the transmission coefficient
\begin{equation}
t=\left|\frac{B}{A}\right|^2=\frac{4\sin^2(qa)}
{\left| -(P_N)_{11}e^{-iqa}+(P_N)_{21}-(P_N)_{12}+(P_N)_{22}e^{iqa}\right|^2}
\label{eq:tc}
\end{equation}
where $(P_N)_{ij}$, $i, j = 1, 2$ are the elements of matrix $P_N$ in 
Eq. (\ref{eq:matrix}).

If we let the atom mass of the  uniform harmonic chain 
be equal to the atom mass in the FK chain, i.e. $m_0=m$, as did in the usual 
study of electron systems\cite{li86}, then we could not 
obtain the transmission coefficients of the phonons with frequency 
larger than $\frac{2}{\sqrt{m}}$. However, for $V\neq 0$, there do exists 
phonons 
whose frequencies are larger than $\frac{2}{\sqrt{m}}$. Therefore, in 
order to study the 
transmission coefficients of {\it all} 
eigenfrequencies, we should let the mass $m_0$ of the uniform harmonic 
chain differ from the mass of the FK chain.
In Figs. \ref{fig1} (a)-(c), we show the transmission coefficients of 
the finite FK chains as the functions of frequency for different $V$.
The parameter 
$\frac{a}{2\pi}=\frac{\sqrt{5}-1}{2}$ is approximated by a converging 
series of truncated fraction: $F_n/F_{n+1}$ ($n=1,2\cdots\cdots$), where
$\{F_n$\} is the Fibonacci sequence. The results of Figs. \ref{fig1}(a)-(c) 
are obtained 
for $N=F_{16}=1597$, $m_0=0.8$, and $m=1$. In our calculations, we first 
obtained the $N$ atomic positions of the equilibrium ground state of the FK 
chain  with free boundary condition, i.e. 
$x_0 = 0$ and $x_{N} = 2\pi Na$, by the 
gradient method\cite{au831,pe83,au832}.
Evidently, for the plane wave with frequency in the gaps of the phonon 
spectra, the transmission coefficients are zero. 
From Fig. (\ref{fig1}), we see that for small  $V$ there exists 
a wider 
frequency range with nonzero transmission coefficients than that for 
large $V$. This can be understood by the following facts.  
For $V=0$, 
there exists only one frequency band from $\omega=0$ to $\frac{2}{\sqrt{m}}$.
When $V\neq 0$, the ground state positions of atoms deviate from that 
in the FK chain without external potential.
For small $V$ ($< V_c$), the ground state positions of atoms are 
periodic or quasiperiodic. 
Therefore, one band splits into several subbands and 
the bandgaps show up. As $V$ increases, more  and more subbands and gaps 
show up.
Also, we found that the range of the eigenfrequency becomes smaller as $V$ 
is increased beyond $V_c$. This implies that the eigenfrequencies are 
attracted somehow for $V>V_c$ (see Fig. 1(c)), which is similar to the band 
gaps of the 
Harper model at the critical point\cite{ge95}. Because the eigenstates of 
the Harper model at critical point are critical, this result is one 
of the signature that the eigenstates of an incommensurate FK chain in 
Cantorus regime are critical.

It is well known\cite{au831} that the phase transition by breaking of
analyticity is manifested by the phonon gap for $V>V_c$. In fact, from 
Fig. 1, we have seen that as $V$ increased to $1.6$ there is a wide range 
of frequencies around $\omega=0$, in which the transmission coefficient is 
zero. This is the direct consequence of the appearance of the phonon gap.
In order to see this 
transition of an incommensurate FK 
model, we calculate the transmission coefficient of a low frequency 
wave at different $V$, which are shown in 
Fig. (\ref{fig2}). In this figure we plot the transmission coefficients of
the plane wave with 
frequency $\omega=10^{-6}$ as a function of $V$ for
the FK segments having different lengths. It is
obviously seen that there is a sharp decrease after $V>V_c$ and 
the decrease becomes sharper for larger $N$. The corresponding particle
number are 1597, 2584, and 4181 for curve 1, 2, and 3, respectively. 
This clearly demonstrates the existence of the phonon gap for $V>V_c$.
In turn, our results illustrated here suggest that the 
transmission coefficient might be a 
very good quantity for probing the existence of phonon gap in the underlying 
system.
Therefore, measuring the transmission coefficient would enable us to 
detect the {\it phase transition by breaking analyticity} 
experimentally.

\section{phonon localization}
\label{sec:three}
The above discussed wave transmission is a macroscopic phenomenon. To some 
extent, it reflects the microscopic origin, namely, the phonon 
excitation in the underlying 
systems. In this section, we would like to study the phonon localization 
from different approaches.

{\bf Participation ratio} --Burkov 
{\sl et. al.} \cite{bu96} studied this quantity for the finite FK chain by 
numerically solving Eq. (\ref{eq:phon}) and computing
the participation ratio (PR)
\begin{equation}
\mbox{PR}=\frac{1}{N}\frac{(\sum_n \psi_n^2)^2}{\sum_n \psi_n^4},
\end{equation}
here $N$ is number of atoms. If the eigenfunction is extended, PR tends 
to be a 
finite limit as $N\rightarrow\infty$. If an eigenfunction is exponentially 
localized, $N\times$PR will be a finite number as the length $N$ goes to 
infinity. Burkov {\sl et al} found that for $V<V_c$, all 
eigenfunctions are extended for $N$ up to $987$ and for $V>V_c$ the PR 
of states near the band edges decrease with increasing $N$, but the true 
exponential localization was not observed. Their method has been limited 
by computer memory (RAM). Therefore, in addition to study the participation 
ratio, we shall also study other quantities such as the transmission 
coefficients and the Thouless exponent. Our results given in 
this section can be regarded as an extension and 
supplement to the previous study by others\cite{bu96,ke97}. 

{\bf Transmission coefficient} -- The transmission coefficient is also a 
very good quantity reflecting
the localization property of eigenstates of electron, and has been used 
widely to study the eigenstates of electron moving in 
random and aperiodic fields\cite{li86}. If the transmission coefficient 
associated 
with the eigenstate tends to be a finite limit as $N\rightarrow\infty$, the 
eigenstate is extended. If the eigenstate is exponentially 
localized, the transmission coefficient will decrease exponentially as 
$N$ increasing. This 
method has an advantage that it can compute transmission coefficients for 
much larger systems.
 
{\bf Thouless exponent} -- Aother important quantity that describes the 
localization of 
eigenstates is the Thouless exponent. The Thouless exponent for an 
eigenfunction corresponding to the eigenfrequency $\omega_i$ of Eq. 
(\ref{eq:phon}) is: 
\begin{equation}
\gamma(\omega_i)=\frac{1}{N-1}\sum_{j\neq i}^{N}\ln 
\left| \omega_j^2-\omega_i^2\right|.
\end{equation}
If $\gamma(\omega_i)$ goes to zero as $N\rightarrow \infty$, then the 
eigenfunction corresponding to $\omega_i$ is localized exponentially.  

In Figs. (\ref{fig3})-(\ref{fig5}), we show the transmission 
coefficients, the Thouless exponents and 
PR's for the eigenstates of a finite FK chain with $N=987$ for $V=0.4$, $1.0$
and $1.6$, respectively.
From the numerical results, we find that the transmission coefficients 
of the phonon eigenstates at the band edges (actually, these are quasi band 
edges because the phonon spectra are the Cantor-like sets, the quasi 
bands consist of many subbands if we consider a larger FK chain) are 
smaller than that in the center of the bands. Because the transmission 
coefficient depends on  $m_0$, its absolute value is not 
meaningful 
for the study of localization of phonon eigenstates. In order to see if 
these eigenstates at band edges are exponentially localized. We calculate 
the transmission coefficients for several FK chains having different 
lengths. Fig. \ref{fig6} shows numerical results for some eigenstates at 
band edges.  We can see that these 
eigenstates are not exponentially localized. This is nicely 
demonstrated by the Thouless exponents shown in Figs. 
\ref{fig3}(b), \ref{fig4}(b) and \ref{fig5}(b). There
the Thouless exponent is about $0.005\sim 0.02$, thus the decay length 
$\xi=1/\gamma\sim 10^2$, which is about the order of the size of the FK 
chains. We also calculated these quantities for 
larger systems (up to 4181, in fact the transfer matrix method allows us 
to go to even larger size, e.g. more than 10,946), but no significant 
difference has been found.

Recently, by using the renormalization group transformation, Ketoja and 
Satija\cite{ke97} have studied the eigenfunctions of the minimum frequency 
for $V>V_c$. They found that phonon eigenstates defy localization and 
remain critical. Furthermore, there exists an infinite sequences of 
parameter values 
in the regime of $V>V_c$ where the renormalization limit cycle degenerates 
into a trivial fixed point\cite{ke97}.

Here, we would like to get a further picture about the localization for 
other different scenarios. We shall study not only the transmission 
coefficients of the degenerated point, but also those cases 
corresponding to the nondegenerated and pseudo-degenerated situation at the 
regime of $V > V_c$.
The results are shown in Fig. (\ref{fig7}). There the
transmission coefficients as functions of $N$ are drawn for 
$V=1.75656208382674\cdots$, $2.33$, and $3.89474285492986\cdots$, respectively. 
These
three values correspond to
degenerated, nondegenerated and pseudo-degenerated cases, 
respectively. We find that there are not qualitative
differences of transmission coefficients for the degenerated parameter 
values and other values of 
$V>V_c$. The only minor difference is that the curves for degenerated and 
pseudo-degenerated 
points are more regular than that for non-degenerated. Another thing 
worthy to be mentioned is that for $N$ fixed, the transmission coefficients 
decreases as $n^{-2}$ 
for $n<\frac{N}{2}$ and increases as $n^2$ for $n>\frac{N}{2}$ [See the 
insets of Fig. \ref{fig7} (a)-(c)].
This  can be understood by realizing that for small part of finite FK 
chain the positions of atoms looks like irrelative, but not real random, 
the plane wave seems propagate through a pseudo-random medium. 
But for large part of finite FK chain, there exists a certain
correlation between 
the positions of atoms, and the correlation increases as $n\rightarrow N$.
The plane wave transports through  a correlated media, and the transmission 
coefficients increase as $N$ increased.

\section{heat conduction}
\label{sec:four}
The properties of the phonon excitations will be also manifested in 
another macroscopic quantity-- the heat conduction. It is time now to  
discuss the transport of energy flux or the heat conduction in the finite 
FK chains. This is a quite interesting problem that attracted much 
attention in recent years\cite{ca84,pr92,le97,hu98,fi98,ph98,al98}. In this 
section, we consider a different model 
of heat conduction, which was discussed by  Rubin  and Greer\cite{ru71} 
originally. In this model,
the chain of $N$ particles (which constitutes the systems) is connected 
at both ends to semi-infinite chains of identical particles. The 
left- and right-ends are put in thermal equilibrium at temperatures 
$T_L$ and $T_R$, respectively. The original Rubin-Greer model was for 
periodic and random mass chains with harmonic nearest neighbor interaction. 
They found that the 
stationary heat flux as a function of $N$ can be expressed in terms of 
the transmission coefficient: 
\begin{equation}
J(N)=\frac{T_L-T_R}{4\pi} \int_{0}^{2}t_N^2(\omega)d\omega
\label{RG}
\end{equation}
It is shown that $J(N)\sim N$ for the uniform and periodic chains and
$J(N)\sim N^{\frac{1}{2}}$ for random mass chains. For the FK 
chains both $T_L$ and $T_R$ must be very low so that the formula (\ref{RG}) 
can be applied. Fig. 
(\ref{fig8}) shows some typical results of $j=J(N)/(T_L-T_R)$ for different 
$V$. In the 
calculation of transmission coefficients, we let the mass of atoms in 
the left- and right-hand semi-infinite uniform harmonic chains be equal to 
the mass of atoms in the FK chain.
We found that the $J(N)\sim N$ for $V<V_c$ and $J(N)\sim N^{\alpha}$
for $V>V_c$. For $V>V_c$, the $\alpha$ depends on the $V$. But its range
is approximately from $0.83$ to $0.87$, which is larger than $0.5$ 
for random systems.  This also implies that the phonon eigenstates of the FK 
chain are extended 
for $V<V_c$ and critical for $V>V_c$. 

It should be pointed out that to express the heat flux in terms of the 
transmission coefficients is valid only at very low temperature, namely 
the particles oscillate nearby their equilibrium positions. In 
fact, this is a linearization result of the FK model. In real case, the FK 
model is a nonlinear, thus for the general simulation of heat 
conduction, one should take the approach of molecular dynamics simulation 
\cite{hu98,fi98}.

\section{discussion and conclusions}
\label{sec:five}
In this paper, we have studied the transmission coefficient of a plane 
wave through the FK chain by making use of the transfer matrix method. We 
have been able to calculate the transmission coefficients of 
{\it all} phonon frequencies. The localization properties of  the phonons 
based on 
the transmission coefficient and Thouless exponents agree with that 
by the participation ratio. 
We also studied the Rubin-Greer-like 
model for the FK 
chain and find that the FK chain likes a periodic chain for $V<V_c$, 
whereas it looks like a chain somewhat between random and periodic 
for $V>V_c$. Our numerical results confirm 
that all eigenstates are critical in Cantorus regime.
This result can be understood as the following. For $V<V_c=0.9716354\cdots$, 
the ground 
state configuration of atoms is quasiperiodic. The $V \cos(x_n^0)$ in 
Eq. (\ref{eq:phon}) is also quasiperiodic and continuous. It corresponds to  
the phonon problem of Harper equation in the extended regime ($V<2$). 
Therefore, all phonon eigenstates are extended. For $V>V_c$, 
the ground state configuration of atoms is a Cantor-like set.
The $V\cos(x_n^0)$ takes only some finite values. Consequently, the 
exponentially
localized state does not exist, and all eigenstates are critical\cite{de86}.

\section*{Acknowledgments}
 
We would like to thank Professor Gang Hu and the members at 
Centre 
for Nonlinear Studies of Hong Kong Baptist University for valuable 
discussions. This work was supported in part by the grants from the Hong 
Kong 
Research Grants Council (RGC) and the Hong Kong Baptist University Faculty 
Research Grant (FRG). Tong's work was also supported in part by 
Natural 
Science Foundation of Jiangsu Province and Natural Science Foundation of 
Jiangsu Education Committee, PR China.

\begin{figure}
\caption{The transmission coefficients as functions of frequency $\omega$ for
different $V$. (a) $V=0.4$, (b) $V=1$, and (c) $V=1.6$. The parameters 
in the
calculations are $N=1597, m_0=0.8$ and $\frac{a}{2\pi} = 
\frac{\sqrt{5}-1}{2}$. The critical value $V_{c}=0.9716543....$.} 
\label{fig1}
\end{figure}

\begin{figure} 
\caption{The transmission
coefficient of low frequency ($\omega=10^{-6}$) wave 
through the FK chain as a function of $V$. For all three curves $m_0=1$ and 
$\frac{a}{2\pi}=\frac{\sqrt{5}-1}{2}$.
The curve 1, 2, and 3 are for $N$=1597, 2548 and 4181,
respectively.}
\label{fig2}
\end{figure}

\begin{figure}
\caption{The transmission coefficients (a), the Thouless exponents $\gamma$ 
(b) 
and  the participation ratio (c) for eigenfunctions of the FK chain at 
$V=0.4$. 
The particle number of the FK chain is $N=987$, the mass of the atom is 
$m_0=0.8$, and the winding number is $\frac{a}{2\pi}=\frac{\sqrt{5}-1}{2}$.} 
\label{fig3} 
\end{figure}

\begin{figure}
\caption{Same as Fig. (\ref{fig3}) but $V=1.0$}
\label{fig4} 
\end{figure}

\begin{figure}
\caption{Same as Fig. (\ref{fig3}) but $V=1.6$
}
\label{fig5} 
\end{figure}

\begin{figure}
\caption{The transmission coefficients of plane waves with frequencies 
near the band edges of phonon spectra as functions of $N$. The solid circles, 
squares and 
diamonds correspond to $\omega=0.369915599017013597$, $0.782580479119728611$
and $0.950165600964576584$ for $V=1.0$, respectively. The solid stars, 
triangles and crosses correspond to $\omega=0.907882097032394975$, 
$1.04046216308972528$ and $2.03331741557249979$ for $V=1.6$, respectively.
In all cases $m_0=0.8$. The lines are drawn for guiding the eye.}
\label{fig6}
\end{figure}

\begin{figure}
\caption{The transmission coefficient at the minimum frequency 
$\omega_{min}$ versus $N$ for different $V$. (a)
$V=1.75656208382674\cdots$,
(b) $V=2.33$, and (c) $V=3.89474285492986\cdots$.
Inserts are $ln(t)$ versus $ln(N)$.}
\label{fig7}
\end{figure}

\begin{figure}
\caption{The ratio ($j=J(N)/(T_R-T_L)$) of the stationary energy flux
to temperature difference for the FK chain as a function of $N$ for 
different $V$. The solid circles, squares, diamonds, and 
triangles are for different values of $V=$0.4, 1.6, 3.0 and 5.0, 
respectively. The lines are drawn for guiding the eye.}
\label{fig8}
\end{figure}

\end{document}